\documentclass[authoryear,12pt,letter,3p]{jowarticle}
\usepackage{graphicx}
\usepackage{amsmath}
\usepackage{amssymb}
\usepackage{natbib}
\usepackage{setspace}
\usepackage{enumerate}
\usepackage[all]{xy}
\usepackage{appendix}
\makeatletter
\renewcommand\section{\@startsection{section}{1}{\z@}{-3.25ex plus -1ex minus -.2ex}{1.5ex plus .2ex}{\normalsize\bf}}
\renewcommand\subsection{\@startsection{subsection}{2}{\z@}{-3.25ex plus -1ex minus -.2ex}{1.5ex plus .2ex}{\normalsize\bf}}
\renewcommand\subsubsection{\@startsection{subsubsection}{3}{\z@}{-3.25ex plus -1ex minus -.2ex}{1.5ex plus .2ex}{\normalsize\bf}}
\makeatother

\newtheorem{thm}{Theorem}[section]

\newtheorem{lem}[thm]{Lemma}
\newtheorem{prop}[thm]{Proposition}

\newtheorem{con}{Criterion}
\newtheorem{conprime}{Criterion}

\numberwithin{equation}{section}

\begin{document}
\begin{frontmatter}
\title{
Are Newtonian Gravitation and Geometrized Newtonian Gravitation Theoretically Equivalent?}
\author{James Owen Weatherall}\ead{weatherj@uci.edu}
\address{Department of Logic and Philosophy of Science\\ University of California, Irvine, CA 92697}
\begin{abstract}I argue that a criterion of theoretical equivalence due to Clark Glymour [\emph{No\^us} \textbf{11}(3), 227--251 (1977)] does not capture an important sense in which two theories may be equivalent.  I then motivate and state an alternative criterion that does capture the sense of equivalence I have in mind.  The principal claim of the paper is that relative to this second criterion, the answer to the question posed in the title is ``yes'', at least on one natural understanding of Newtonian gravitation.\end{abstract}
\end{frontmatter}
\doublespacing

\section{Introduction}

Are Newtonian gravitation and geometrized Newtonian gravitation (Newton-Cartan Theory) equivalent theories?  Clark \citet{GlymourTETR, GlymourEG, GlymourTE} has articulated a natural criterion of theoretical equivalence and argued that, by this criterion, the answer is ``no''.\footnote{Glymour's criterion has recently been a topic of debate on other grounds: see, for instance, \citet{Halvorson,HalvorsonReply}, \cite{GlymourReply}, and \citet{Coffey}.}  I will argue here that the situation is more subtle than Glymour suggests, by characterizing a robust sense in which two theories may be equivalent that Glymour's criterion does not capture.  This alternative sense of equivalence, which is in the same spirit as Glymour's, is best construed as a friendly amendment.\footnote{For a detailed discussion of the relationship between the present proposal and other senses of equivalence in the literature, including Glymour's cirterion, see \citet{Barrett+Halvorson}.  For an overview of applications of the present criterion, see \citet{WeatherallLandry}.}  Still, it will turn out that by this alternative criterion, Newtonian gravitation \emph{is} equivalent to geometrized Newtonian gravitation---at least on one way of construing Newtonian gravitation.\footnote{David \citet{Zaret} has also replied to Glymour on this question.  But his argument is markedly different than the one presented here, and \citet{Glymour+Spirtes} offer what I take to be an effective reply.}  It follows that there exist realistic theories that are equivalent in a robust and precise sense, but which apparently disagree regarding certain basic features of the world, such as whether spacetime is curved.

The paper will proceed as follows.  I will begin by briefly reviewing the two versions of Newtonian gravitation.  I will then describe Glymour's criterion for theoretical equivalence, according to which the two versions of Newtonian gravitation fail to be equivalent.  Next, I will apply Glymour's criterion to two formulations of electromagnetism that, I will argue, should be (and typically are) taken to be equivalent.  It will turn out that these theories fail to be equivalent by Glymour's criterion of equivalence.  In the following sections, I will develop an alternative notion of equivalence between theories that I will argue does capture the sense in which these two formulations of electromagnetism are equivalent.  I will then return to the question of principal interest in the present paper, arguing that there are two ways of construing standard (nongeometrized) Newtonian gravitation.  I will state and prove a simple proposition to the effect that, by the alternative criterion, on one of the two ways of construing standard Newtonian gravitation (but not the other), it is theoretically equivalent to geometrized Newtonian gravitation.  I will conclude by drawing some morals concerning the interpretation of physical theories.  Proofs of selected propositions appear in an appendix.

\section{Two formulations of Newtonian gravitation}

The two theories with which I am principally concerned are Newtonian gravitation (NG) and a variant of Newtonian gravitation due to \'Elie \citet{Cartan1,Cartan2} and Kurt \citet{Friedrichs}, called ``Newton-Cartan theory'' or ``geometrized Newtonian gravitation'' (GNG).\footnote{For background on geometrized Newtonian gravitation, see \citet{MalamentGR} or \citet{Trautman}.}  In NG, gravitation is a force exerted by massive bodies on other massive bodies.  It is mediated by a gravitational potential, and in the presence of a (non-constant) gravitational potential, massive bodies will accelerate.  In GNG, meanwhile, gravitation is ``geometrized'' in much the same way as in general relativity: the geometrical properties of spacetime depend on the distribution of matter, and conversely, gravitational effects are manifestations of this geometry.  Despite these differences, however, there is a precise sense, which I will state below, in which the theories are empirically equivalent.  The central question of the paper is whether they are also equivalent in some stronger sense.

On both theories, spacetime is represented by a four dimensional manifold of spacetime events, which I will assume throughout is $\mathbb{R}^4$. This manifold is equipped with two (degenerate) metrics: a temporal metric $t_{ab}$ of signature $(1,0,0,0)$ that assigns temporal lengths to vectors, and a spatial metric $h^{ab}$ of signature $(0,1,1,1)$ that (indirectly) assigns spatial lengths to vectors.\footnote{Throughout the paper I use the abstract index notation, explained in \citet[\S 1.4]{MalamentGR}.}  These are required to satisfy $h^{ab}t_{bc}=\mathbf{0}$ everywhere.  There always exists (at least locally) a covector field $t_a$ such that $t_{ab}=t_at_b$; a spacetime is \emph{temporally orientable} if this field can be defined globally.  In what follows, I will limit attention to temporally orientable spacetimes.  Finally, spacetime is endowed with a derivative operator $\nabla$ that is compatible with both metrics, in the sense that $\nabla_a t_{b}=\mathbf{0}$ and $\nabla_a h^{bc}=\mathbf{0}$ everywhere.  Since $\nabla_a t_b=\mathbf{0}$ and $\mathbb{R}^4$ is simply connected, there exists a globally defined smooth function $t:M\rightarrow \mathbb{R}$ such that $t_a=\nabla_a t$.  This function allows us to foliate spacetime into maximal $t=const$ hypersurfaces, each with a positive definite metric induced by $h^{ab}$.  These surfaces represent space at various times; here we assume that each of these hypersurfaces is diffeomorphic to $\mathbb{R}^3$ and complete relative to the metric induced by $h^{ab}$.

With these assumptions, the four elements just described define a \emph{classical spacetime}, written $(M,t_a,h^{ab},\nabla)$.  Matter in both theories is represented by its mass density field, which is a smooth scalar field $\rho$.  Massive point particles are represented by their worldlines---smooth curves whose tangent vector fields $\xi^a$ satisfy $\xi^at_a\neq0$.  Such curves are called \emph{timelike}.

In this context, NG is the theory whose models are classical spacetimes with flat ($R^a{}_{bcd}=\mathbf{0}$) derivative operators, endowed with a gravitational potential, which is a scalar field $\varphi$ satisfying Poisson's equation, $\nabla_a\nabla^a\varphi=4\pi\rho$.\footnote{Here $\nabla^a\varphi=h^{ab}\nabla_b\varphi.$}  A massive point particle whose worldline has tangent field $\xi^a$ will accelerate according to $\xi^n\nabla_n\xi^a=-\nabla^a\varphi$.  In the geometrized version of the theory, meanwhile, the derivative operator is permitted to be curved and the gravitational potential is omitted.  The curvature field associated with the derivative operator satisfies a geometrized version of Poisson's equation, $R_{ab}=4\pi\rho t_a t_b$, and in the absence of any external (i.e., non-gravitational) interactions, massive particles traverse timelike geodesics of this curved derivative operator.  In both cases, we take the ``empirical content'' of the theory to consist in the allowed trajectories of massive bodies, in the absence of any non-gravitational force, given a particular mass density.

Given a model of NG, it is always possible to produce a (unique) model of GNG that agrees on empirical content in this sense.
\begin{prop}[\citet{Trautman}]\label{geometrization}Let $(M,t_a,h^{ab},\overset{f}{\nabla})$ be a flat classical spacetime, let $\varphi$ and $\rho$ be smooth scalar fields satisfying Poisson's equation with respect to $\overset{f}{\nabla}$, and let $\overset{g}{\nabla}=(\overset{f}{\nabla},C^a_{\;\;bc})$, with $C^a{}_{bc}=-t_bt_c\overset{f}{\nabla}\,^a\varphi$.\footnote{The notation $\nabla'=(\nabla,C^a{}_{bc})$ is explained in \citet[Prop. 1.7.3]{MalamentGR}.  Briefly, the action of the derivative operator $\nabla'$ on any tensor field can be expressed as the sum of the action of $\nabla$ on that field and terms involving a ``connecting field'' $C^a{}_{bc}$.  Specifying $\nabla$ and $C^a{}_{bc}$ is thus sufficient to define $\nabla'$.}  Then $(M,t_a,h^{ab},\overset{g}{\nabla})$ is a classical spacetime; $\overset{g}{\nabla}$ is the unique derivative operator on $M$ such that given any timelike curve with tangent vector field $\xi^a$, $\xi^n\overset{g}{\nabla}_n\xi^a=\mathbf{0}$ iff $\xi^n\overset{f}{\nabla}_n\xi^a=-\overset{f}{\nabla}\,^a\varphi$; and the Riemann curvature tensor relative to $\overset{g}{\nabla}$, $\overset{g}{R}\,^a_{\;\;bcd}$, satisfies (1) $\overset{g}{R}_{ab}=4\pi\rho t_a t_b$, (2) $\overset{g}{R}{}^a_{\;\;b}{}^c_{\;\;d}=\overset{g}{R}{}^{c}_{\;\;d}{}^a_{\;\;b}$, and (3) $\overset{g}{R}{}^{ab}{}_{cd}=\mathbf{0}$.\end{prop}
\noindent It is also possible to go in the other direction, as follows.
\begin{prop}[\citet{Trautman}]\label{recovery} Let $(M, t_a,h^{ab},\overset{g}{\nabla})$ be a classical spacetime that satisfies conditions (1)-(3) in Prop. \ref{geometrization} for some smooth scalar field $\rho$.  Then there exists a smooth scalar field $\varphi$ and a flat derivative operator $\overset{f}{\nabla}$ such that $(M,t_a,h^{ab},\overset{f}{\nabla})$ is a classical spacetime; given any timelike curve with tangent vector field $\xi^a$, $\xi^n\overset{g}{\nabla}_n\xi^a=\mathbf{0}$ iff $\xi^n\overset{f}{\nabla}_n\xi^a=-\overset{f}{\nabla}\,^a\varphi$; and $\varphi$ and $\rho$ together satisfy Poisson's equation relative to $\overset{f}{\nabla}$.\end{prop}

It is important emphasize that the pair $(\overset{f}{\nabla},\varphi)$ in Prop. \ref{recovery} is not unique.  A second pair $(\overset{f}{\nabla}{}',\varphi')$ will satisfy the same conditions provided that (1) $\overset{g}{\nabla}{}^a\overset{g}{\nabla}{}^b(\varphi'-\varphi)=\mathbf{0}$ and (2) $\overset{f}{\nabla}{}'=(\overset{f}{\nabla},C^a{}_{bc})$, with $C^a{}_{bc}=t_bt_c\overset{g}{\nabla}{}^a(\varphi'-\varphi)$.  Note, too, that Prop. \ref{recovery} holds only if conditions (1)-(3) from Prop. \ref{geometrization} are satisfied.  The geometrized Poisson equation, condition (1), has already been assumed to hold of models of GNG; for present purposes, I will limit attention to models of GNG that also satisfy conditions (2) and (3).\footnote{\label{gravField1} Note that throughout this section, one could substitute ``gravitation field'' for ``gravitational potential'' by replacing every instance of $\nabla^a\varphi$ with a smooth vector field $\varphi^a$ satisfying $\nabla^{[a}\varphi^{b]}=\mathbf{0}$.  The choice makes no difference to the results below, though some readers may think a theory committed to a gravitational field is more plausible than one committed to a gravitational potential.}

\section{Glymour on theoretical equivalence}

I will now turn to Glymour's account of theoretical equivalence.  The underlying intuition is that two theories are theoretically equivalent if (1) they are empirically equivalent and (2) they are mutually inter-translatable.\footnote{Glymour does not state that empirical equivalence is a necessary condition for theoretical equivalence, though he does appear to take theoretical equivalence to be strictly stronger than empirical equivalence, and, as \citet{SklarSN} emphasizes, empirical equivalence is a substantive interpretive constraint that goes beyond any formal relations between two theories.}  In general, empirical equivalence is a slippery concept, but we will not discuss it further.  For present purposes, it suffices to stipulate that the theories being compared \emph{are} empirically equivalent, in the precise senses described.

The idea behind the second condition, of mutual inter-translatability, is that two theories should be said to be equivalent if they have precisely the same expressive resources, or in other words, if anything one can say about the world in one theory can be said equally well in the other, and vice versa. Glymour makes this criterion precise via the notion of definitional equivalence in first order logic.\footnote{For details on explicit definability and definitional equivalence, see \citet[Ch. 2.6]{Hodges}. See also the classic work by \citet{Bouvere1,Bouvere2}, and more recently, \citet{Barrett+Halvorson}.} Suppose that $L$ and $L^+$ are first-order signatures, with $L\subseteq L^+$.  An explicit definition of a symbol in $L^+$ in terms of $L$ is a sentence in $L^+$ that asserts the equivalence between that symbol (appropriately used) and some formula in $L$.  Given a theory $T$ in $L$, by appending explicit definitions of the symbols in $L^+/L$ to $T$, we may extend $T$ to a theory in $L^+$.  The resulting theory is a \emph{definitional extension of $T$ in $L^+$}.  Now suppose $T_1$ and $T_2$ are first-order theories in signatures $L_1$ and $L_2$, respectively, with $L_1\cap L_2=\emptyset$.  Then $T_1$ and $T_2$ are \emph{definitionally equivalent} if and only if there are first order theories $T_1^+$ and $T_2^+$ in $L_1\cup L_2$ such that $T_1^+$ is a definitional extension of $T_1$; $T_2^+$ is a definitional extension of $T_2$; and $T_1^+$ and $T_2^+$ are logically equivalent.  Definitional equivalence captures a sense of inter-translatability in that, given any pair of definitionally equivalent theories $T_1$ and $T_2$ and a formula $\varrho$ in the language of $T_1$, it is always possible to translate $\varrho$ into a formula in the language of $T_2$, and then back into a formula in the language of $T_1$ that is $T_1-$provably equivalent to $\varrho$.\footnote{For more on this sort of translation, see \citet{Barrett+HalvorsonQuine}.}

Definitional equivalence is a natural notion of equivalence for first order theories.  But it is difficult to apply directly to physical theories, since we rarely have first order formulations available.  For this reason, Glymour works with a model-theoretic variant of definitional equivalence.  Suppose $T_1$ and $T_2$ are definitionally equivalent theories, and suppose that $A_1$ is a model of $T_1$.  Then it is always possible to expand $A_1$ into a model $A$ of $T_1^+$, the definitional extension of $T_1$.  Since $T_1^+$ and $T_2^+$ (the extension of $T_2$) are logically equivalent, $A$ is also a model of $T_2^+$.  We may thus turn $A$ into a model $A_2$ of $T_2$ by restricting $A$ to symbols in the language of $T_2$.  The whole process can then be reversed to recover $A_1$.  In this sense, definitionally equivalent theories ``have the same models'' insofar as a model of one theory can be systematically transformed into a model of the other theory, and vice versa.\footnote{It is essential that one can go from a model $A_1$ of $T_1$ to a model $A_2$ of $T_2$, and then back to the \emph{same} model $A_1$ of $T_1$.  See \citet{Hungarians}.}

Using this model-theoretic characterization of definitional equivalence as inspiration, Glymour proposes the following criterion of equivalence for physical theories expressed in terms of covariant objects on a manifold.\footnote{Actually, all Glymour claims is that clause (2) of this criterion is a necessary condition for theoretical equivalence.  I am extrapolating when I say that the two clauses together are also sufficient.}
\begin{con}\label{Glymour}
Theories $T_1$ and $T_2$ are theoretically equivalent if for every model $M_1$ in $T_1$, there exists a unique model $M_2$ in $T_2$ that (1) has the same empirical content as $M_1$ and (2) is such that the geometrical objects associated with $M_2$ are uniquely and covariantly definable in terms of the elements of $M_1$ and the geometrical objects associated with $M_1$ are uniquely and covariantly definable in terms of $M_2$, and vice versa.
\end{con}
GNG and NG fail to meet this criterion. The reason is that, as noted at the end of the last section, models of NG are not uniquely determined by models of GNG.

\section{A problem case for Glymour?}

I will presently argue that criterion \ref{Glymour} does not capture an important sense in which two physical theories may be equivalent.  I will do so by displaying two ``theories'' (actually, formulations of a single theory) that usually are (I claim correctly) taken to be equivalent, but which fail to meet Glymour's criterion.  These theories correspond to two ways of presenting classical electromagnetism on Minkowski spacetime, $(M,\eta_{ab})$.\footnote{Minkowski spacetime is a (fixed) relativistic spacetime $(M,\eta_{ab})$ where $M$ is $\mathbb{R}^4$, $\eta_{ab}$ is a flat Lorentzian metric, and the spacetime is geodesically complete.  For more on these two formulations of electromagnetism, see \citet{WeatherallGauge}.}

On the first formulation of the theory, which I will call EM$_1$, the dynamical variable a smooth, antisymmetric tensor field $F_{ab}$ on $M$.  This field is called the Faraday tensor; it represents the electromagnetic field on spacetime. The Faraday tensor satisfies Maxwell's equations, which may be written as (1) $\nabla_{[a}F_{bc]}=\mathbf{0}$ and (2) $\nabla_a F^{ab}=J^b$, where $J^b$ is a smooth vector field representing charge-current density.  (Here $\nabla$ is the Levi-Civita derivative operator compatible with $\eta_{ab}$.)  Models on this formulation may be written $(M,\eta_{ab},F_{ab})$.\footnote{Here and in what follows, we do not include the charge-current density in specifications of models of electromagnetism, as this field can be uniquely reconstructed from the other fields, given Maxwell's equations.}  On the second formulation, which I will call EM$_2$, the dynamical variable is a smooth vector field $A_a$ on $M$, called the $4-$vector potential.  This field satisfies the differential equation $\nabla_a\nabla^a A^b - \nabla^b\nabla_a A^a = J^b$.  Models may be written $(M,\eta_{ab},A_a)$.

These two formulations are systemically related.  Given a vector potential $A_a$ on $M$, one may define a Faraday tensor by $F_{ab}=\nabla_{[a}A_{b]}$.  This tensor will satisfy Maxwell's equations for some $J^a$ if $A_a$ satisfies the differential equation above for the same $J^a$.  Conversely, given a Faraday tensor $F_{ab}$ satisfying Maxwell's equations (for some $J^a$), there always exists a vector potential $A_a$ satisfying the required differential equation (for that $J^a$), such that $F_{ab}=\nabla_{[a}A_{b]}$. We stipulate that on both formulations, the empirical content of a model is exhausted by its associated Faraday tensor.  In this sense, the theories are empirically equivalent, since for any model of EM$_1$, there is a corresponding model of EM$_2$ with the same empirical content (for some fixed $J^a$), and vice versa.

But are EM$_1$ and EM$_2$ equivalent by Glymour's criterion?  No.  Given any model $(M,\eta_{ab},A_a)$ of EM$_2$, I can uniquely determine a model $(M,\eta_{ab},F_{ab})$ of EM$_1$ by taking $F_{ab}=\nabla_{[a}A_{b]}$.  But given a model $(M,\eta_{ab},F_{ab})$ of EM$_1$, there are generally many corresponding models of EM$_2$.  In particular, if $F_{ab}=\nabla_{[a}A_{b]}$ for some 4-vector potential $A_a$, then $F_{ab}=\nabla_{[a}\tilde{A}_{b]}$ will also hold if (and only if) $\tilde{A}_a=A_a+G_a$, where $G_a$ is a closed one form (i.e., $\nabla_{[a}G_{b]}=\mathbf{0}$).  Thus uniqueness fails in the EM$_2$ to EM$_1$ direction.

What should one make of this result? On the one hand, Glymour's criterion seems to capture something important: the failure of uniqueness suggests that EM$_2$ distinguishes physical situations that EM$_1$ cannot distinguish.  On the other hand, EM$_1$ and EM$_2$ are usually taken to be different formulations of the same theory; they are intended to have precisely the same theoretical content. The tension concerns the relationship between the models of EM$_2$.  The transformations between models of EM$_2$ associated with the same Faraday tensor are often called ``gauge transformations''.  On their standard interpretation, models related by a gauge transformation are \emph{physically equivalent}, in the sense that they have the capacity to represent precisely the same physical situations.\footnote{The status of the vector potential arguably changes in quantum mechanics. See \citet{BelotEM}.}  Thus EM$_2$ does \emph{not} distinguish situations that EM$_1$ cannot.  And indeed, it seems to me that there is a clear and robust sense in which two theories should be understood as equivalent if, on their standard interpretations, they differ only with regard to features that, by the lights of the theories themselves, have no physical content.

\section{An alternative criterion}

Thus far, I have introduced a criterion of theoretical equivalence and argued that it fails to capture the sense in which EM$_1$ and EM$_2$ are equivalent.  In the present section, I will present a criterion of equivalence that does capture the sense in which EM$_1$ and EM$_2$ are equivalent.  To motivate this new criterion, note first that there are actually two reasons that EM$_1$ and EM$_2$ fail to meet Glymour's criterion.   The first problem concerns the failure of a model of EM$_1$ to correspond to a unique model of EM$_2$.  In particular, if we want a sense of theoretical equivalence that captures the sense in which EM$_1$ and EM$_2$ are equivalent to one another, we need to be able to accommodate the possibility that not all of the structure of models of EM$_2$ is salient.  That is, we want a sense of unique recovery \emph{up to physical equivalence}.

One way to make this idea precise is to modify the definition of models of EM$_2$.  Instead of characterizing a model as a triple $(M,\eta_{ab},A_a)$, we might take a model to be a triple $(M,\eta_{ab},[A_a])$, where $[A_a]$ is the equivalence class of physically equivalent vector potentials, $[A_a]=\{\tilde{A}_a : \tilde{A}_a = A_a + G_a\text{ for closed }G_a\}$.  This approach explicitly equivocates between physically equivalent vector potentials.  Call the theory whose models are so characterized EM$'_2$.
\begin{prop}\label{EMunique} For any model $(M,\eta_{ab},F_{ab})$ of EM$_1$, there is a unique model $(M,\eta_{ab},[A_a])$ of EM$_2'$ such that $F_{ab}=d_aX_a$ for every $X_a\in [A_a]$.\end{prop}
\noindent Thus we \emph{do} have unique recovery of models of EM$_2'$ from models of EM$_1$.

We still face a second problem, however.  This problem concerns what is meant by ``covariant definability''.\footnote{I am particularly grateful to Thomas Barrett and Jeff Schatz for discussions about and suggestions on this paragraph and the next two.  But they should not be held responsible for what I say!}  Glymour does not make this notion precise, and it is not obvious that there is a unique or particularly natural way to do so that would meet all of the desiderata one might impose on a notion of ``definability''.  To see the difficulty,  observe that even in first order logic, a distinction is made between ``explicit definition'' and ``implicit definition'', and there are a number of substantive and subtle theorems that show that these different notions of definability are equivalent.\footnote{The classic results here are Beth's theorem and Svenonius' theorem.  See \citet[Theorem 6.64 \& Corollary 10.5.2]{Hodges} and the surrounding discussion.}  These theorems do not hold in many other logics of interest---including second order logic, where implicit and explicit definability come apart in general.\footnote{For a survey of definability properties in various logics, including second order logic, see \citet{Makowsky+Shelah}; see also \citet{Craig}, \citet{Gostanian+Hrbacek}, and, for a more accessible treatment, \citet{Andreka+NemetiNotes}.}  

The theories we are working with here are not first order theories---or at least, we have not specified a first order ``language of electromagnetism'' or ``language of geometrized Newtonian gravitation,'' nor have we presented axioms of any sort or relied only on proofs in a first order system.  Indeed, it is not clear that satisfactory first order theories of electromagnetism or geometrized Newtonian gravitation exist, and more importantly, the question of whether such theories \emph{do} exist does not seem to arise in practice.  This suggests that there are, at least in principal, different notions of definability available for these theories, none of which has been made precise in the present context.  So it is hard to know how to proceed.\footnote{For instance, if I do not know what the ``language of electromagnetism'' is, I can hardly write down an explicit definition in that language!}

To be sure, this issue does not really arise in Glymour's own treatment of NG and GNG, or in the relationship between EM$_1$ and EM$_2$ as discussed in the previous section.  The reason is that the failures of uniqueness in both cases show that, \emph{whatever} one means by covariant definability, one will not be able to produce the necessary definitions.  But the problem becomes acute once we move to EM$'_2$, where Prop. \ref{EMunique} guarantees that we \emph{do} have unique recovery.  It is just not clear what the further requirement of ``covariant definability'' amounts to.

To address this problem, let us return to Glymour's original argument.  Recall that the basic strategy was to adapt definitional equivalence to a setting more conducive to analyzing physical theories by identifying a model theoretic consequence of definitional equivalence.  Now, though, we see that Glymour's criterion is not as well suited to evaluating physical theories as it appeared to be.  But we need not abandon the basic strategy, of looking to relations between collections of the models of definitionally equivalent theories.  One such relation that seem particularly attractive concerns not bare sets of models, as in Glymour's condition, but rather \emph{categories} of models associated with definitionally equivalent theories.\footnote{A \emph{category} consists of (1) a collection of objects $A,B,C\ldots$; (2) a collection of arrows $f,g,h\ldots$; and (3) assignments to each arrow $f$ of a pair of objects, $dom(f)$ and $cod(f)$, called the domain and codomain the arrow, respectively.  (We abbreviate this by $f:dom(f)\rightarrow cod(f)$.)  We require that for any arrows $f,g$ such that $cod(f)=dom(g)$, there exists an arrow $g\circ f : dom(f)\rightarrow cod(g)$ called the \emph{composition of $f$ and $g$}; and for any object $A$, there exists an arrow $1_A: A\rightarrow A$ called the \emph{identity arrow}.  Together, these must satisfy: (1) for any arrows $f,g,h$,  if $(h\circ g)\circ f$ exists, then $(h\circ g)\circ f = h\circ(g\circ f)$; and (2) for any arrow $f:A\rightarrow B$, $f\circ 1_A = f = 1_B\circ f$.  The category of models of a theory $T$ has models of $T$ as objects and elementary embeddings as arrows.  For more on categories and the related notions described below, see \citet{MacLane}, \citet{Borceux}, or \citet{Leinster}, among many other excellent texts.  For more on the present proposal for understanding theoretical equivalence using category theory, see \citet{Halvorson}, \citet{HalvorsonOxford}, \citet{Barrett+Halvorson}, and \citet{WeatherallLandry}.}  In particular, if theories $T$ and $T'$ are definitionally equivalent, then their \emph{categories} of models are \emph{isomorphic}.\footnote{A \emph{functor} $F:\mathbf{C}\rightarrow\mathbf{D}$ is a map between categories that takes objects to objects and arrows to arrows, and which preserves identity arrows and composition.  Given functors $F:\mathbf{C}\rightarrow\mathbf{D}$ and $G:\mathbf{D}\rightarrow \mathbf{E}$, the composition $G\circ F$, defined  in the obvious way, is always a functor.  A functor $F:\mathbf{C}\rightarrow\mathbf{D}$ is an \emph{isomorphism} of categories if there is a functor $F^{-1}:\mathbf{D}\rightarrow\mathbf{C}$ such that $F\circ F^{-1}=1_{\mathbf{D}}$ and $F^{-1}\circ F=1_{\mathbf{C}}$, where $1_{\mathbf{C}}:\mathbf{C}\rightarrow\mathbf{C}$ and $1_{\mathbf{D}}:\mathbf{D}\rightarrow\mathbf{D}$ are functors that act as the identity on objects and arrows.  The result cited in the text is proved by \citet{Barrett+Halvorson}.  (The result they state concerns \emph{equivalence} of categories, to be discussed below, but in fact they show the stronger thing as well.)}

This observation suggests the following alternative criterion of equivalence.
\begin{conprime}\label{CatGlymour}
Two theories are theoretically equivalent just in case there exists an isomorphism between their categories of models that preserves empirical content.
\end{conprime}
I call this criterion \ref{CatGlymour} because it bears a very close relationship to Glymour's original criterion.  For one, criterion \ref{CatGlymour} is motivated by the same basic intuition about inter-translation as criterion \ref{Glymour}: in both cases, the basic idea is that two theories are equivalent if I can take whatever one theory says about the world and translate it, in some appropriate sense, into the other theory, and vice versa, in a way that loses nothing.  We have even adopted both the same starting point for making this idea of ``mutual intertranslatability'' precise---definitional equivalence---and the same strategy for adapting it to the present context, of moving to models of theories.

This is not to say that the resulting criteria are the same.  In fact, even in first order logic, isomorphism between categories of models is strictly weaker than definitional equivalence.\footnote{Again, this is discussed in full detail in \citet{Barrett+Halvorson}.  It is not known how much weaker categorical isomorphism is than definitional equivalence, or Morita equivalence, which is a weakening of definitional equivalence that allows one to define new sorts.  Note that the model theoretic criterion Glymour begins with is actually equivalent to definitional equivalence \citep{Bouvere1}, at least in simple cases, though that is of little comfort if, as I have argued, it cannot actually be applied in realistic cases. See \citet{GlymourReply} and \citet{HalvorsonReply} for a recent discussion of these issues.}  In the present context, one might characterize the difference as follows: Glymour's criterion attempts to spell out ``inter-translatability'' using some combination of ``semantic'' considerations---translating directly between models of the theory---and ``syntactic'' ones, insofar as his criterion requires some notion of ``definition''.  The present criterion, meanwhile, drops the definability requirement, but adds the requirement that further structure be preserved by the maps relating the models of the theories---namely, the category theoretic structure encoding information about automorphisms and other elementary embeddings of models.  One may think of this as capturing the idea that the models of the two theories have the same structure---and thus, have the capacity to represent the same physical situations.

As hoped, this new criterion is readily applied to physical theories. To do so in the present case, we define a category $\mathbf{EM}_1$ whose objects are models of EM$_1$ and whose arrows are isometries of Minkowski spacetime that preserve the Faraday tensor, and a category $\mathbf{EM}_2'$ whose objects are models of EM$_2'$ and whose arrows are isometries of Minkowski spacetime that preserve the equivalence classes of vector potentials.\footnote{What is meant by ``preserve the equivalence classes'' is described in more detail in Lemma \ref{EMdefinable}, below.} Given these categories, we may then prove the following result.
\begin{prop}\label{EMisomorphic}There exists an isomorphism of categories between $\mathbf{EM}_1$ and $\mathbf{EM}'_2$ that preserves empirical content.\end{prop}
Prop. \ref{EMisomorphic} shows that there \emph{is} a sense in which EM$_1$ and EM$'_2$ are equivalent---namely, the sense given by criterion \ref{CatGlymour}.

To get a clearer sense of what is going on, and how Prop. \ref{EMisomorphic} relates to criterion \ref{Glymour}, observe that in the course of proving Prop. \ref{EMisomorphic}, one establishes the following.
\begin{lem}\label{EMdefinable} Let $(M,\eta_{ab},F_{ab})$ and $(M,\eta_{ab},F'_{ab})$ be models of EM$_1$ and let $(M,\eta_{ab},[A_a])$ and $(M,\eta_{ab},[A_a]')$ be the unique corresponding models of EM$_2'$. Then an isometry $\chi:M\rightarrow M$ is such that $\chi_*(F_{ab})=F'_{ab}$ iff $[\chi_*(A_a)]=[A_a]'$.\footnote{Here $\chi_*$ is the pushforward along $\chi$, defined for differential forms because $\chi$ is a diffeomorphism.}\end{lem}
This result provides a sense in which the models of EM$_2'$ might be said to be \emph{implicitly} definable from the models of EM$_1$: any map that preserves a Faraday tensor, as well as the other structure of Minkowski spacetime, automatically preserves the equivalence class of vector potentials associated with that Faraday tensor, and vice versa.  One might even think of this result as establishing a perfectly good sense in which models of EM$'_2$ \emph{are} (implicitly) covariantly definable from models of EM$_1$ after all.

So much for EM$_1$ and EM$_2'$.  But what about EM$_2$, the alternative formulation of electromagnetism we began with?  After all, it was \emph{this} theory that we originally wanted to claim was equivalent to EM$_1$.  We may define a category of models of this theory, too: as a first pass, we take $\mathbf{EM}_2$ to be the category whose objects are models of EM$_2$ and whose arrows are isometries of Minkowski spacetime that preserve the vector potential.  But this category is \emph{not} isomorphic to $\mathbf{EM}_1$---and so, on this representation of EM$_2$, EM$_1$ and EM$_2$ are still not equivalent, even by criterion \ref{CatGlymour}.  The problem is the same as with Glymour's criterion: there is a failure of unique recovery.

We have already argued that this sort of non-uniqueness is spurious, at least on the standard interpretation of EM$_2$, because models related by a gauge transformation should be counted as physically equivalent.  The category $\mathbf{EM}_2$ does not reflect this equivalence between models, because in general, two models that differ by a gauge transformation will not be isomorphic in this category.  On the other hand, we also know that there is another class of mapping between models that \emph{does} reflect this sort of physical equivalence---namely, the gauge transformations themselves.  These maps do not appear as arrows in the category $\mathbf{EM}_2$, which suggests that if we want to represent EM$_2$ accurately, in the sense of representing it in a way that accords with what structure we take to be physically significant on the standard interpretation, we need a different category, one that includes information about the gauge transformations.

We define such a category as follows: we take $\overline{\mathbf{EM}}_2$ to be the category whose objects are models of EM$_2$ and whose arrows are pairs of the form $(\chi,G_a):(M,\eta_{ab},A_a)\rightarrow (M,\eta_{ab},A'_a)$, where $G_a$ is closed and $\chi$ is an isometry that preserves the (gauge transformed) vector potential $A_a+G_a$, in the sense that $\chi^*(A'_a)=A_a + G_a$.
\begin{prop}\label{EMcategory}$\overline{\mathbf{EM}}_2$ is a category.\end{prop}
Note that $\overline{\mathbf{EM}}_2$ is naturally understood to include the arrows of $\mathbf{EM}_1$, which may be identified with pairs of the form $(\chi,0)$, the gauge transformations, which are arrows of the form $(1_M,G_a)$, and compositions of these.

Intuitively speaking, $\overline{\mathbf{EM}}_2$ is the result of taking $\mathbf{EM}_2$ and ``adding'' arrows corresponding to the gauge transformations.  Simply adding arrows in this way, however, does not yield a category that is (empirical-content-preservingly) isomorphic to $\mathbf{EM}_1$.  The reason is that the extra arrows do not address the failure of unique recovery.  But that does not mean this exercise was in vain.  Although there is not an \emph{isomorphism} between $\mathbf{EM}_1$ and $\overline{\mathbf{EM}}_2$ that preserves empirical content, there is an \emph{equivalence} of categories that does so.\footnote{An \emph{equivalence of categories} is a pair of functors $F:\mathbf{C}\rightarrow\mathbf{D}$ and $G:\mathbf{D}\rightarrow\mathbf{C}$ that are \emph{almost inverses} in the sense that given any object $A$ of $\mathbf{C}$, there is an isomophism $\eta_A:A\rightarrow G\circ F(A)$, where these isomorphisms collectively satisfy the requirement that for any arrow $f:A\rightarrow B$ of $\mathbf{C}$, $\eta_B\circ f = G\circ F(f)\circ \eta_A$; and likewise, \emph{mutatis mutandis}, for any object of $\mathbf{D}$.}
\begin{prop}\label{EMequivalent} There is an equivalence of categories between $\mathbf{EM}_1$ and $\overline{\mathbf{EM}}_2$ that preserves empirical content.\end{prop}
Equivalent categories may be thought of as categories that are isomorphic ``up to object isomorphism''---which is precisely the notion of equivalence we argued we were looking for between EM$_1$ and EM$_2$ at the end of the last section.

The considerations in the last paragraph suggest a new, still weaker criterion.
\begin{con}\label{CatGauge} Two theories are theoretically equivalent just in case there exists an equivalence between their categories of models that preserves empirical content.\end{con}
Prop. \ref{EMequivalent} establishes that EM$_1$ and EM$_2$ \emph{are} theoretically equivalent by this new criterion---so long as we represent EM$_2$ by $\overline{\mathbf{EM}}_2$, rather than $\mathbf{EM}_2$.  It is in this sense, I claim, that the two formulations of electromagnetism should be taken to be equivalent.

What can be said about this sense of equivalence?  In fact, the same interpretation can be given for criterion \ref{CatGauge} as for criterion \ref{CatGlymour}.  Once again, we are capturing a sense in which models of one theory can be ``translated'' into models of another theory, and then back, without losing any information---or in other words, the models of the two theories have the same structure, and one can map between models of the two theories without losing that structure.  The difference between \ref{CatGlymour} and \ref{CatGauge} comes down to whether we require the ``translation'' to be unique, or merely unique up to isomorphism.  But if our goal is to capture the idea that the models of the two theories have the same amount of structure, then it is hard to see why we would want more than uniqueness up to isomorphism, since after all, isomorphic models have the same structure, qua models of the theory in question.\footnote{More generally, \citet{Barrett+Halvorson} show that if two theories are \emph{Morita} equivalent, which is similar to definitional equivalence, but with the flexibility to define new sorts, then their categories of models are equivalent, but \emph{not} necessarily isomorphic.  So there is reason to think that even in the first order case, we should be interested in categorical equivalence, rather than isomorphism.}

As a final remark, let me observe that criterion \ref{CatGauge} also captures the sense in which EM$_2$ and EM$_2'$ are equivalent.  In particular, these theories are \emph{not} equivalent by criterion \ref{CatGlymour}, even though $\mathbf{EM}_2'$ and $\overline{\mathbf{EM}}_2$ may seem to be equally good ways of capturing ``gauge equivalence'' in a formal representation of EM$_2$.  This reflects the more general fact that for most mathematical purposes, equivalence of categories is a more natural and fruitful notion of ``sameness'' of categories than isomorphism.  It also suggests that criterion \ref{CatGlymour} is at best an awkward half-way point once we have begun thinking in the present terms.


\section{Are NG and GNG theoretically equivalent?}

With these new criteria in hand, we now return to the question at the heart of the paper.  To apply either criterion to NG and GNG, however, one first needs to say what categories we will use to represent the theories.  For GNG, there is a clear choice. We represent GNG by the category $\mathbf{GNG}$ whose objects are classical spacetimes $(M,t_a,h^{ab},\nabla)$ satisfying the required curvature conditions from Prop. \ref{geometrization}, and whose arrows are diffeomorphisms that preserve the classical metrics and the derivative operator.\footnote{Given a diffeomorphism $\chi:M\rightarrow M'$ and derivative operators $\nabla$ and $\nabla'$ on $M$ and $M'$ respectively, we say that $\chi$ preserves $\nabla$ if for any tensor field $\lambda^{a_1\cdots a_r}_{b_1\cdots b_s}$ on $M$, $\chi_*(\nabla_n\lambda^{a_1\cdots a_r}_{b_1\cdots b_s})=\nabla'_n\chi_*(\lambda^{a_1\cdots a_r}_{b_1\cdots b_s})$.}

NG is more complicated, however.  There is a natural option for the objects: they are classical spacetimes with gravitational potentials $(M,t_a,h^{ab},\nabla,\varphi)$, where $\nabla$ flat.  But we face a choice concerning the arrows, corresponding to a choice about which models of NG are physically equivalent.
\begin{enumerate}[{Option} 1.]
\item One takes models of NG that differ with regard to the gravitational potential to be distinct.\footnote{\label{gravField2} Alternatively, one could replace ``gravitational potential'' with ``gravitational field'' to yield a distinct, and perhaps more plausible, option.  (Recall footnote \ref{gravField1}.)  But the difference does not matter for the present discussion.} \label{nonGauge}
\item One takes models of NG whose gravitational potential and derivative operators are related by the transformation $\varphi\mapsto \varphi'=\varphi+\psi$ and $\nabla\mapsto\nabla'=(\nabla,t_b t_c\nabla^a\psi)$, for any smooth $\psi$ satisfying $\nabla^a\nabla^b\psi=\mathbf{0}$, to be equivalent.\footnote{Note that, since \emph{all} of the derivative operators considered in NG and GNG agree once one raises their index, one can characterize the gauge transformation with regard to any of them without ambiguity.}\label{Gauge}
\end{enumerate}
In the second case, one takes the gravitational potential to be a gauge quantity, much like the vector potential in electromagnetism.  In the first case, one does not.

These two options suggest different categories.  In particular, we define $\mathbf{NG}_1$ to be the category whose objects are as above, and whose arrows are diffeomorphisms that preserve the classical metrics, the derivative operator, and the gravitational potential, and we define $\mathbf{NG}_2$ to be the category with the same objects, but whose arrows are pairs $(\chi,\psi)$, where $\psi$ is a smooth scalar field satisfying $\nabla^a\nabla^b\psi=\mathbf{0}$, and $\chi$ is a diffeomorphism that preserves the classical metrics and the (gauge transformed) derivative operator $\nabla'=(\nabla,t_b t_c\nabla^a\psi)$ and gravitational potential $\varphi+\psi$.  The first category corresponds to option \ref{nonGauge}, while the second corresponds to option \ref{Gauge}.  Since these options correspond to different interpretations of the formalism, I will treat them as prima facie distinct theories, labeled as NG$_1$ and NG$_2$, respectively, in what follows.

What considerations might lead one to prefer one option over the other?  The first option better reflects how physicists have traditionally thought of Newtonian gravitation.  On the other hand, this option appears to distinguish between models that are not empirically distinguishable, even in principle.  Moreover, there are physical systems for which option \ref{nonGauge} leads to problems, such as cosmological models with homogeneous and isotropic matter distributions, where option \ref{nonGauge} generates contradictions that option \ref{Gauge} avoids.\footnote{For more on this point, see the debate between John \citet{Norton1, Norton2} and David \citet{MalamentNC}.  Arguably, Newton himself recognized the empirical equivalence of models related by these transformations---for instance, see the discussions of Corollary VI to the laws of motion in \citet{DiSalle}; see also \citet{Saunders, Knox2, WeatherallMaxwell}.}  These latter arguments strike me as compelling, and I tend to think that option \ref{Gauge} is preferable.  But I will not argue further for this thesis, and for the purposes of the present paper, I will remain agnostic about these options.

We may now ask: are any of these theories pairwise equivalent by either criterion?  None of these theories are equivalent by criterion \ref{CatGlymour}, effectively for the reason that NG and GNG fail to be equivalent by Glymour's original criterion.\footnote{Note, however, that one could construct an alternative presentation of NG$_2$ analogous to EM$_2'$, in such a way that this \emph{would} be equivalent to GNG by criterion \ref{CatGlymour}.  Moreover, if one restricts attention to the collections of models of NG and GNG in which (1) the matter distribution is supported on a spatially compact region and (2) the gravitational field (for models of NG) vanishes at spatial infinity, then $\mathbf{NG}_1$, $\mathbf{NG}_2$, and $\mathbf{GNG}$ are all equivalent by \emph{both} criteria.}  Moreover, NG$_1$ is not equivalent to either GNG or NG$_2$ by criterion \ref{CatGauge}.  But GNG and NG$_2$ \emph{are} equivalent by criterion \ref{CatGauge}.
\begin{prop}\label{NGequivalent}
There is an equivalence of categories between $\mathbf{NG}_2$ and $\mathbf{GNG}$ that preserves empirical content.
\end{prop}
The situation is summarized by table \ref{equivalences}.

\begin{table}
\centering
\begin{tabular}{| c | c | c |}
\hline
 &Condition \ref{CatGlymour} &Condition \ref{CatGauge}\\
\hline
NG$_1$ and NG$_2$ &Inequivalent &Inequivalent\\
 \hline
NG$_1$ and GNG & Inequivalent &Inequivalent\\
\hline
NG$_2$ and GNG & Inequivalent &\textbf{Equivalent}\\
\hline
\end{tabular}
\caption{A summary of the equivalences and inequivalences of NG and GNG, by the standards set by conditions \ref{CatGlymour} and \ref{CatGauge}.\label{equivalences}}
\end{table}

\section{Interpreting physical theories: some morals}

I have now made the principal arguments of the paper.  In short, criterion \ref{Glymour} does not capture the sense in which EM$_1$ and EM$_2$ are equivalent.  However, there is a natural alternative criterion that does capture the sense in which EM$_1$ and EM$_2$ are equivalent.  And by this criterion, GNG and NG are equivalent too, if one adopts option \ref{Gauge} above.  Moreover, criterion \ref{CatGauge} highlights an important distinction between two ways of understanding NG.

There are a few places where one might object.  One might say that \emph{no} formal criterion captures what it would mean for two theories to be equivalent.\footnote{For versions of this worry, see \citet{SklarSN} and \citet{Coffey}.}  One might also reject the significance of the particular criteria discussed here. I do not agree with these objections, but I will not consider them further.  For the remainder of this paper, I will suppose that criterion \ref{CatGauge} does capture an interesting and robust sense in which these theories may be equivalent.  If this is right, there are several observations to make.

First of all, the arguments here support one of Glymour's principal claims, which is that there exist empirically equivalent, theoretically inequivalent theories.  This is because even if NG$_2$ and GNG are theoretically equivalent, NG$_1$ and GNG are still inequivalent, even by condition \ref{CatGauge}.  Glymour's further claim that GNG is better supported by the empirical evidence, on his account of confirmation, is only slightly affected, in that one needs to specify that GNG is only better supported than NG$_1$.  This makes sense: the reason, on Glymour's account, that GNG is better supported than NG is supposed to be that NG makes additional, unsupported ontological claims regarding the existence of a gravitational potential.\footnote{Once again, one could substitute ``gravitational field'' for ``gravitational potential,'' \emph{mutatis mutandis}.  Recall footnotes \ref{gravField1} and \ref{gravField2}.}  But one can understand the difference between NG$_1$ and NG$_2$ in this way as well, since NG$_2$ explicitly equivocates between models that differ with regard to their gravitational potentials.

There is another purpose to which Glymour puts these arguments, however.  There is a view, originally due to \citet{PoincareSH} and \citet{Reichenbach}, that the geometrical properties of spacetime are a matter of convention because there always exist empirically equivalent theories that differ with regard to (for instance) whether spacetime is curved or flat.\footnote{For a clear and detailed description of the positions that have been defended on the epistemology of geometry in the past, see \citet{SklarSTST}; see also \citet{Weatherall+Manchak}.}  Glymour argues against conventionalism by pointing out that the empirical equivalence of two theories does not imply that they are equally well confirmed, since the theories may be theoretically inequivalent.  But the present discussion suggests that there is another possibility that is not often considered: theories that attribute apparently distinct geometrical properties to the world may be \emph{more} than just empirically equivalent.  They may provide different, but equally good, ways of representing the \emph{same} structure in the world.

As I have just noted, one way of understanding NG$_2$ is as a theory on which the gravitational potential is not a real feature of the world, because the gravitational potential is not preserved by mappings that reflect physical equivalence.  GNG, meanwhile, does not make any reference to a gravitational potential. In this sense GNG and NG$_2$ appear to have the same ontological implications, at least with regard to gravitational potentials.  Still, GNG and NG$_2$ do differ in one important way. In particular, in generic models of GNG, spacetime is curved.  In all models of NG$_2$, meanwhile, \emph{spacetime is flat}.  Now, interpreting this second fact is somewhat subtle.  Since we have taken models of NG$_2$ to be equivalent if they are related by gauge transformations that in general do \emph{not} preserve derivative operators---even though they \emph{do} preserve curvature---it is not correct to say that models of NG$_2$ represent spacetime as flat, since in fact, they do not posit any particular parallel transport properties.  Another way of making this point is to observe that  although in all models of NG$_2$, parallel transport of vectors is path independent, the result of parallel transporting any particular vector along a given (fixed) curve will generally vary even between equivalent models, because the derivative operator varies with gauge transformations.\footnote{I am grateful to Oliver Pooley for pressing this point.}

One might conclude from this that GNG provides a more perspicuous representation of spacetime geometry, since the apparent geometry of the models of NG$_2$ is obscured by the gauge transformations.\footnote{\citet{Knox} makes a closely related point.}  But there is another option available.  As \citet{WeatherallMaxwell} shows, there is some geometrical structure shared by all isomorphic models of NG$_2$, beyond just the metric structure: namely, they all agree on a standard of rotation.  In other words, we may think of models of NG$_2$ as positing enough structure to say when a body (say) is rotating, but not enough to say that it is undergoing unaccelerated (inertial) motion, full stop.\footnote{In other words, one might think of the models of NG$_2$ as having the structure of Maxwell-Huygens spacetime, as in \citet{WeatherallMaxwell}; see also \citet{Saunders} and \citet{Knox2}.}

In any case, one thing seems clear.  Models of GNG represent spacetime as curved, whereas models of NG$_2$ do not.  Thus, at least in this context, there is a sense in which classical spacetime admits equally good, \emph{theoretically} equivalent descriptions as either curved or not.  Let me emphasize that this view is not a recapitulation of traditional conventionalism about geometry.  For one, it is not a general claim about spacetime geometry; the view here depends on the details of the geometry of classical spacetime physics.  Indeed, there is good (though perhaps not dispositive) reason to think that general relativity, for instance, is \emph{not} equivalent to a theory on which spacetime is flat, by any of the criteria discussed here.\footnote{See \citet{Knox} and \citet{Weatherall+Manchak} for evidence supporting this claim.}  More generally, I do not believe that it is a matter of convention whether we choose one empirically equivalent theory over another.  I agree with Glymour that there are often very good reasons to think one theory is better supported than an empirically equivalent alternative.  Rather, the point is that in some cases, apparently different descriptions of the world---such as a description on which spacetime is flat and one on which it is curved---amount to the same thing, insofar as they have exactly the same capacities to represent physical situations.  In a sense, they say the same things about the world.

The suggestion developed in the last paragraph will worry some readers.  Indeed, one might be inclined to reject criterion \ref{CatGauge} (or even criterion \ref{CatGlymour}) on the grounds that one has antecedent or even \emph{a priori} reason for thinking that there is, in all cases, an important distinction---perhaps a \emph{metaphysical} distinction---between a theory that says spacetime is curved and one that does not.  Two theories that disagree in this regard could not both be true, because at most one could accurately reflect the facts about the curvature of spacetime, and thus, two such theories could not be equivalent.  I think this position is probably tenable.  But it seems to me to get things backwards.  At the very least, there is another way of looking at matters, whereby one allows that the distinctions that one can sensibly draw depends on the structure of the world.  And the best guide to understanding what those distinctions are will be to study the properties of and relationships between our best physical theories.

\appendix

\section{Proofs of propositions}

\noindent \textbf{Proof of Prop. \ref{EMunique}.}

\noindent Suppose there were vector potentials $A_a$ and $\tilde{A}_a$ such that $[A_a]\neq[\tilde{A}_a]$, but for every $X_a\in [A_a]$, $\nabla_{[a}X_{b]}=\nabla_{[a}\tilde{A}_{b]}=F_{ab}$.  Then $\nabla_{[a}(X_{b]}-\tilde{A}_{b]})=\mathbf{0}$ for every $X_a\in [A_a]$, and thus $X_a-\tilde{A}_a$ is closed for every $X_a\in[A_a]$.  Thus $[A_a]\subseteq [\tilde{A}_a]$.  A similar argument establishes that $[\tilde{A}_a]\subseteq[A_a]$.\hspace{.25in}$\square$

\noindent \textbf{Proof of Lemma \ref{EMdefinable}.}

\noindent Suppose we have an isometry $\chi$ s.t. $\chi_*(F_{ab})=F'_{ab}$.  Then for every $X_a\in[A_a]$, we have $\chi_*(\nabla_{[a}X_{b]})=\chi_*(F_{ab})=F'_{ab}$.  But exterior derivatives commute with pushforwards along diffeomorphisms, and so $\chi_*(\nabla_{[a}X_{b]})=\nabla_{[a}\chi_*(X_{b]})=F'_{ab}$.  Thus by Prop. \ref{EMunique}, $[\chi_*(A_a)]=[A'_a]$.  Conversely, if $\chi_*(A_a)\in[A_a]$, then $F'_{ab}=\nabla_{[a}\chi_*(A_{b]})=\chi_*(\nabla_{[a}A_{b]}=\chi_*(F_{ab})$.\hspace{.25in}$\square$

\noindent \textbf{Proof of Prop. \ref{EMisomorphic}.}

\noindent The isomorphism is given by $F:EM'_2\rightarrow EM_1$ acting on models as $(M,\eta_{ab},[A_a])\mapsto (M,\eta_{ab},\nabla_{[a}A_{b]})$ and acting on arrows as the identity.  That this yields an isomorphism is an immediate consequence of Prop. \ref{EMunique}, Lemma \ref{EMdefinable}, and basic facts about the composition of pushforward maps.\hspace{.25in}$\square$

\noindent \textbf{Proof of Prop. \ref{EMcategory}.}

\noindent $\mathbf{EM}_2$ includes identity arrows, which are pairs of the form $(1_M,0)$; (2) it contains all compositions of arrows, since given any two arrows $(\chi,G_a)$ and $(\chi',G'_a)$ with appropriate domain and codomain, $(\chi',G_a')\circ (\chi,G_a)=(\chi'\circ\chi,\chi^*(G'_a)+G_a)$ is also an arrow; and (3) composition of arrows is associative, since given three pairs $(\chi,G_a)$, $(\chi',G_a')$, and $(\chi'',G_a'')$ with appropriate domain and codomain, $(\chi'',G_a'')\circ((\chi',G_a')\circ (\chi,G_a))=(\chi'',G_a'')\circ(\chi'\circ\chi,\chi^*(G'_a)+G_a)=(\chi''\circ (\chi'\circ\chi),\chi^*\circ\chi'^*(G_a'') + \chi^*(G_a)'+G_a)=((\chi''\circ \chi')\circ\chi,\chi^*(\chi'^*(G_a'') + G_a')+G_a)=(\chi''\circ \chi',\chi'^*(G_a'') + G_a')\circ(\chi,G_a)=((\chi'',G_a'')\circ(\chi',G_a'))\circ(\chi,G_a)$.\hspace{.25in}$\square$

\noindent \textbf{Proof of Prop. \ref{EMequivalent}.}

\noindent It suffices to show that there is a functor from $\overline{\mathbf{EM}}_2$ to $\mathbf{EM}_1$ that is full, faithful, and essentially surjective, and which preserves $F_{ab}$.  Consider the functor $E:\overline{\mathbf{EM}}_2\rightarrow\mathbf{EM}_1$ defined as follows: $E$ acts on objects as $(M,\eta_{ab},A_a)\mapsto (M,\eta_{ab},\nabla_{[a}A_{b]})$ and on arrows as $(\chi,G_a)\mapsto\chi$.  This functor clearly preserves $F_{ab}$. It is also essentially surjective, since given any $F_{ab}$, there always exists some $A_a$ such that $\nabla_{[a}A_{b]}=F_{ab}$.  Finally, to show that it is full and faithful, we need to show that for any two objects $(M,\eta_{ab},A_a)$ and $(M,\eta_{ab},A'_a)$, the induced map on arrows between these models is bijective.  First, suppose there exist two distinct arrows $(\chi,G_a),(\chi',G'_a):(M,\eta_{ab},A_a)\rightarrow (M,\eta_{ab},A'_a)$.   If $\chi\neq\chi'$ we are finished, so suppose for contradiction that $\chi=\chi'$.  Since by hypothesis these are distinct arrows, it must be that $G_a\neq G'_a$.  But then $A_a+G_a\neq A_a + G'_a$, and so $\chi_*(A_a+G_a)\neq\chi_*(A_a+G'_a)$.  So we have a contradiction, and $\chi\neq\chi'$.  Thus the induced map on arrows is injective.  Now consider an arrow $\chi:E((M,\eta_{ab},A_a))\rightarrow E((M,\eta_{ab},A'_a))$.  This is an isometry such that $\chi_*(\nabla_{[a}A_{b]})=\nabla_{[a}A'_{b]}$.  It follows that $\chi_*(\nabla_{[a}A_{b]}-\nabla_{[a}\chi^*(A'_{b]}))=\mathbf{0}$, and thus that $\nabla_{[a}A_{b]}-\nabla_{[a}\chi^*(A'_{b]})$ is closed.  So there is an arrow $(\chi,\chi^*(A'_a)-A_a):(M,\eta_{ab},A_a)\rightarrow(M,\eta_{ab},A'_a)$ such that $E((\chi,\chi^*(A'_a)-A_a))=\chi$, and the induced map on arrows is surjective.\hspace{.25in}$\square$

\noindent  \textbf{Proof of Prop. \ref{NGequivalent}.}

\noindent This argument follows the proof of Prop. \ref{EMequivalent} closely.  Consider the functor $E:\mathbf{NG}_2\rightarrow\mathbf{NG}_1$ defined as follows: $E$ takes objects to their geometrizations, as in Prop. \ref{geometrization}, and it acts on arrows as $(\chi,\psi)\mapsto\chi$.  This functor preserves empirical content because the geometrization lemma does; meanwhile, Prop. \ref{recovery} ensures that the functor is essentially surjective.  We now show it is full and faithful.  Consider any two objects $A=(M,t_a,h^{ab},\nabla,\varphi)$ and $A'=(M',t_a',h'^{ab},\nabla',\varphi')$.  Suppose there exist distinct arrows $(\chi,\psi),(\chi',\psi'):A\rightarrow A'$, and suppose (for contradiction) that $\chi=\chi'$.  Then $\psi\neq\psi'$, since the arrows were assumed to be distinct.  But then $\varphi+\psi\neq\varphi+\psi'$, and so $(\varphi+\psi)\circ\chi\neq(\varphi+\psi')\circ\chi$. Thus $\chi\neq\chi'$ and $E$ is faithful.  Now consider any arrow $\chi:E(A)\rightarrow E(A')$; we need to show that there is an arrow from $A$ to $A'$ that $E$ maps to $\chi$.  I claim that the pair $(\chi,\varphi'\circ\chi-\varphi):A\rightarrow A'$ is such an arrow.  Clearly if this arrow exists in $\mathbf{NG}_2$, $E$ maps it to $\chi$, so it only remains to show that this arrow exists.  First, observe that since $\chi$ is an arrow from $E(A)$ to $E(A')$, $\chi:M\rightarrow M'$ is a diffeomorphism such that $\chi_*(t_a)=t'_a$ and $\chi_*(h^{ab})=h'^{ab}$.  Moreover, $\chi_*(\varphi + (\varphi'\circ\chi-\varphi))=\chi_*(\varphi'\circ\chi) = \varphi'\circ(\chi\circ\chi^{-1})=\varphi'$, so $\chi$ maps the gauge transformed potential associated with $A$ to the potential associated with $A'$.  Now consider the action of $\chi$ on the derivative operator $\nabla$.  We need to show that for any tensor field $\lambda^{a_1\cdots a_r}_{b_1\cdots b_s}$, $\chi_*(\tilde{\nabla}_n\lambda^{a_1\cdots a_r}_{b_1\cdots b_s})=\nabla'_n\chi_*(\lambda^{a_1\cdots a_r}_{b_1\cdots b_s})$, where $\tilde{\nabla}=(\nabla,t_bt_c\nabla^a(\varphi'\circ\chi-\varphi))$ is the gauge transformed derivative operator associated with $A$.  We will do this for an arbitrary vector field; the argument for general tensor fields proceeds identically.  Consider some vector field $\xi^a$.  Then $\chi_*(\tilde{\nabla}_n\xi^a)=\chi_*(\nabla_n\xi^a - t_n t_m\xi^m\nabla^a(\varphi'\circ\chi-\varphi))=\chi_*(\overset{g}{\nabla}_n\xi^a-t_nt_m\xi^m\nabla^a\varphi- t_n t_m\xi^m\nabla^a(\varphi'\circ\chi-\varphi))=\chi_*(\overset{g}{\nabla}_n\xi^a)-\chi_*(t_n t_m\xi^m\nabla^a(\varphi'\circ\chi))$, where $\overset{g}{\nabla}=(\nabla,t_bt_c\nabla^a\varphi)$ is the derivative operator associated with $E(A)$.  Now, we know that $\chi:E(A)\rightarrow E(A')$ is an arrow of $\mathbf{GNG}$, so $\chi_*(\overset{g}{\nabla}_n\xi^a)=\overset{g}{\nabla}_n\chi_*(\xi^a)$.  Moreover, note that the definitions of the relevant $C^a{}_{bc}$ fields guarantee that $\nabla^a\lambda^{a_1\cdots a_r}_{b_1\cdots b_s}=\overset{g}{\nabla}{}^a(\lambda^{a_1\cdots a_r}_{b_1\cdots b_s})$ and similarly for $\nabla'$ and $\overset{g}{\nabla}{}'$.  Thus we have $\chi_*(\overset{g}{\nabla}_n\xi^a)-\chi_*(t_n t_m\xi^m\nabla^a(\varphi'\circ\chi))=\overset{g}{\nabla}{}_n'\chi_*(\xi^a)-t'_nt'_m\chi_*(\xi^m)\nabla'^a(\varphi'\circ(\chi\circ\chi^{-1}))=\overset{g}{\nabla}{}_n'\chi_*(\xi^a) -t'_nt'_m\chi_*(\xi^m)\nabla'^a\varphi'=\nabla'_n\chi_*(\xi^a)$, where $\overset{g}{\nabla}{}'=(\nabla',-t_bt_c\nabla^a\varphi')$ is the derivative operator associated with $E(A')$.  So $\chi$ does preserve the gauge transformed derivative operator.  The final step is to confirm that $\nabla^a\nabla^b(\varphi'\circ\chi-\varphi)=\mathbf{0}$.  To do this, again consider an arbitrary vector field $\xi^a$ on $M$.  We have just shown that $\nabla'_a\chi_*(\xi^b)-\chi_*(\nabla_a\xi^b)=-\chi_*(t_at_m\xi^m\nabla^b(\varphi'\circ\chi-\varphi)$.  Now consider acting on both sides of this equation with $\nabla'^a$.  Beginning with the left hand side (and recalling that $\nabla$ and $\nabla'$ are both flat), we find: $\nabla'^n\nabla'_a\chi_*(\xi^b)-\nabla'^n\chi_*(\nabla_a\xi^b)=\nabla'_a\overset{g}{\nabla}{}'^n\chi_{*}(\xi^b)-\chi_*(\nabla_a\overset{g}{\nabla}{}^n\xi^b)=\overset{g}{\nabla}{}'_a\overset{g}{\nabla}{}'^n\chi_*(\xi^a)- t'_at'_m(\nabla'^b\varphi')\overset{g}{\nabla}{}'^n\chi_*(\xi^m)-\chi_*(\overset{g}{\nabla}{}_a\overset{g}{\nabla}{}^n\xi^b)+\chi_*(t_at_m(\nabla^b\varphi)\overset{g}{\nabla}{}^n\xi^m)= \chi_*(t_at_m(\nabla^b(\varphi-\varphi'\circ\chi))\overset{g}{\nabla}{}^n\xi^m)$.  The right hand side, meanwhile, yields $-\nabla'^n(\chi_*(t_at_m\xi^m\nabla^b(\varphi'\circ\chi-\varphi))=\chi_*(t_at_m(\nabla^n\xi^m)\nabla^b(\varphi-\varphi'\circ\chi))+\chi_*(t_at_m\xi^m\nabla^n\nabla^b(\varphi-\varphi'\circ\chi))$.  Comparing these, we conclude that $\chi_*(t_at_m\xi^m\nabla^n\nabla^b(\varphi-\varphi'\circ\chi))=\mathbf{0}$, and thus $t_at_m\xi^m\nabla^n\nabla^b(\varphi-\varphi'\circ\chi)=\mathbf{0}$.  But since $t_a$ is non-zero and this must hold for \emph{any} vector field $\xi^a$, it follows that $\nabla^a\nabla^b(\varphi'\circ\chi-\varphi)=\mathbf{0}$.  Thus $E$ is full.\hspace{.25in}$\square$

\section*{Acknowledgments}
This material is based upon work supported by the National Science Foundation under Grant No. 1331126.  Thank you to Steve Awodey, Jeff Barrett, Thomas Barrett, Ben Feintzeig, Sam Fletcher, Clark Glymour, Hans Halvorson, Eleanor Knox, David Malament, John Manchak, Colin McLarty, John Norton, Cailin O'Connor, Oliver Pooley, Sarita Rosenstock, Jeff Schatz, Kyle Stanford, and Noel Swanson for helpful conversations on the topics discussed here, and to audiences at the Southern California Philosophy of Physics Group, the University of Konstanz, and Carnegie Mellon University for comments and discussion.   I am particularly grateful to Erik Curiel, Clark Glymour, Hans Halvorson, David Malament, and two anonymous referees for comments on a previous draft, and to Thomas Barrett for detailed discussion and assistance concerning the relationship between categorical equivalence and definitional equivalence.

\singlespacing
\bibliography{equivalence}
\bibliographystyle{elsarticle-harv}

\end{document}